\documentclass[a4paper,12pt]{article}
\usepackage{color}
\usepackage{jheppub}
\usepackage{amsmath,amssymb,latexsym}
\usepackage{url}
\begin{document}

\hfill {\tt CERN-TH-2016-191}

\hfill {\tt TIFR/TH/16-30}

\title{A Higgs in the Warped Bulk and LHC signals}
\author{F. Mahmoudi$^{a,b,}$\footnote{Also Institut Universitaire de France, 103 boulevard Saint-Michel, 75005 Paris, France},}
\author{U. Maitra$^c$,}
\author{N. Manglani$^{d,e}$,} 
\author{K.Sridhar$^c$}
\affiliation[a]{Univ Lyon, Univ Lyon 1, ENS de Lyon, CNRS, Centre de Recherche Astrophysique de Lyon UMR5574, F-69230 Saint-Genis-Laval, France}
\affiliation[b]{Theoretical Physics Department, CERN, CH-1211 Geneva 23, Switzerland}
\affiliation[c]{Department of Theoretical Physics, Tata Institute of Fundamental Research, Homi Bhabha Road, Colaba, Mumbai 400 005, India}
\affiliation[d]{Department of Physics, University of Mumbai,Kalina, Mumbai 400098, India}
\affiliation[e]{Shah and Anchor Kutchhi Engineering College, Mumbai 400088, India}

\emailAdd{nazila@cern.ch}
\emailAdd{ushoshi@theory.tifr.res.in}
\emailAdd{namratam@physics.mu.ac.in}
\emailAdd{sridhar@theory.tifr.res.in}

\abstract{Warped models with the Higgs in the bulk can generate light Kaluza-Klein (KK) Higgs modes consistent with the electroweak precision analysis. The first KK mode of the Higgs ($h_{1}$)
could lie in the 1-2 TeV range in the models with a bulk custodial symmetry. We find that the $h_{1}$ is gaugephobic and decays dominantly into a $t\bar{t}$ pair.
We also discuss the search strategy for $h_{1}$ decaying to $t\bar{t}$ at the Large Hadron Collider. We used substructure tools to suppress the large QCD background associated with this channel.
We find that $h_{1}$ can be probed at the LHC run-2 with an integrated luminosity of 300 fb$^{-1}$.}

\maketitle

%%%%%%%%%%%%%%%%%%%%%%%%%%%%%%%%%%%%%%%%%%%%%%%%%%%%%%%%%%%%%%%%%%%%%%%%%%%%%

\section{Introduction}
\noindent The Randall-Sundrum model (RS model) \cite{Randall:1999ee},
as originally proposed,  is a five-dimensional model with a
warped metric
\begin{equation}
	ds^2=e^{-2 A(y)}\eta_{\mu\nu}dx^{\mu}dx^{\nu}- dy^2 \;,
	\label{metric}
\end{equation}
with the fifth dimension $y$ compactified on an $S^1/Z^2$ orbifold of 
radius, $R$.  Two branes are located at  $y=0$ and $y=\pi R \equiv L$ 
and are called the UV and the IR branes respectively. 

Starting with a bulk gravity action one can show that the solutions to
the Einstein equation imply for the warp factor $A(y)$
\begin{equation}
	A(y) = \pm k \vert y \vert\;,
	\label{e6.11}
\end{equation}
where $k^2 \equiv - \Lambda /12 M^3$ with $M$ being the Planck
scale. 
A value of $kL \sim 30$ is sufficient, through the warp factor,
to generate a factor of $v / M \sim 10^{-16}$ 
(where $v$ is the vacuum expectation value of the SM Higgs field) 
thereby stabilising the gauge hierarchy. This suppression factor
is, however, material for all fields localised on the IR brane and,
indeed, in the original RS model this was the case for all SM fields
with only gravity localised in the bulk. With SM fields localised on
the brane, mass scales which suppress dangerous
higher-dimensional operators responsible for proton decay or neutrino
masses also become small and this spells a disaster for the RS model. 

Wisdom gleaned from AdS/CFT correspondence also gives an understanding
of the need to go beyond the original RS model. The fields localised
on the IR brane turn out, through the correspondence, to be composites
of operators in the four-dimensional field theory that is dual to the
RS model. The latter then turns out to be dual to a theory where all
the SM fields are composite, which is not viable. However, a theory
of partial compositeness is viable and can survive experimental
constraints. This corresponds to a RS model where the SM fields
are localised in the bulk.

This was, in fact, the motivation to move the SM fields into the
bulk and construct what are called the Bulk RS models. 
For reviews, see Refs.~\cite{Gherghetta:2010cj, Raychaudhuri:2016}.
In such models,
often, the Higgs is still kept localised on the IR brane so that
the gauge-hierarchy solution discussed above continues to hold.
The big gain that accrues in the Bulk RS models
is that the differential localisation of SM fermions in the bulk 
gives rise in a natural way to the Yukawa-coupling hierarchy
\cite{Pomarol:1999ad, Gherghetta:2000qt, Grossman:1999ra}. 
The other features of Bulk RS models are that they give rise to small mixing 
angles in the Cabibbo-Kobayashi-Maskawa (CKM) matrix, provide a 
natural way of obtaining gauge-coupling universality and allow for the 
suppression of flavour-changing neutral currents
\cite{Burdman:2003nt, Huber:2003tu, Casagrande:2008hr, Bauer:2009cf, 
Agashe:2004cp}.

As shown in later work on Bulk RS models \footnote{For a review, see 
\cite{Quiros:2013yaa}.}, even the Higgs need not be
sharply localised on the IR brane but only somewhere close to it in order
to address gauge-hierarchy. This freedom allows for more interesting
model-building possibilities. It is this latter class of models
which will be the focus of the present paper.

The serious issue to contend with in Bulk RS models is that of
electroweak precision. In models with only gauge bosons propagating
in the bulk, the constraints on the masses of the Kaluza-Klein (KK) gauge
bosons are very strong (of the order of 25 TeV) though this is somewhat 
ameliorated by also
allowing SM fermions in the bulk, especially with fermions of the
first and second generation localised close to the UV brane. Even
in this case, there are unacceptably large couplings of the KK gauge
bosons to the Higgs resulting in severe $T$-parameter constraints. 
One way of addressing this problem is called the 
Custodial symmetry model. In this model we have an enlarged gauge symmetry 
\cite{Agashe:2003zs, Agashe:2006at} 
in the bulk i.e an $SU(3)_c \times  SU(2)_L \times SU(2)_R \times
U(1)_y$ that acts like the custodial symmetry of the
SM in protecting the $\rho$ parameter and this extended group is then 
broken on the IR brane to recover the SM gauge group. This extended
symmetry takes care of the $T$-parameter but non-oblique $Z \rightarrow b
\bar b$ corrections,
coming from the fact that the fermions are not all localised at the same
point in the bulk, persist which are then addressed by a suitable
choice of fermion transformations under the custodial symmetry group.
The bound on the lightest
KK gauge boson mode comes down to about 3 TeV \cite{Davoudiasl:2009cd, Iyer:2015ywa}. 
\footnote{There are other approaches in dealing with the electroweak precision constraints such as the deformed metric model 
\cite{Cabrer:2010si,Cabrer:2011fb} or a model using brane kinetic term \cite{Carena:2003fx}, but we will not consider these approaches here.}

The upshot of the above discussion is that, it is
possible to get the masses of the KK modes of SM particles within
the reach of collider searches. Indeed, there is already
a significant amount of literature suggesting search strategies for KK gauge
bosons \cite{Agashe:2006hk, Lillie:2007yh, Guchait:2007jd, Allanach:2009vz,
Agashe:2007ki, Agashe:2008jb, Iyer:2016yjb} and 
KK fermions \cite{Agashe:2004ci,Davoudiasl:2007wf} at the Large Hadron Collider (LHC). In contrast, KK modes
of bulk Higgs have not received their due attention. 
The zero mode of the bulk Higgs has been studied in \cite{Davoudiasl:2005uu, Cacciapaglia:2006mz}
and in \cite{Frank:2016vtv} the CP-odd excitation 
of the bulk Higgs in the deformed metric model has been studied.
It is to the search for the first KK excitation of the Higgs in the context of the custodial symmetric model at the LHC that we devote the rest of this paper.

\section{Bulk Higgs Models}

The action with the Higgs propagating in the bulk~\cite{Quiros:2013yaa} is given by
\begin{equation}
 S = \int{d^{4}x} dy \sqrt{-g} (D_{M} \Phi D^{M} \Phi - m^{2}\Phi^{\dagger}\Phi + 2\sum_{j=0,1}(-1)^{j}\lambda^{j}(\Phi) \delta(y - y_{j}) + L_{yuk})
 \label{action_5}\;,
\end{equation}
where $y_{0}~=~0,~y_{1}~=~\pi R$,\\
$$ -\lambda^{1}(\Phi) = -\frac{M_{1}}{k}|\Phi^{\dagger}\Phi| + 2 \frac{\gamma}{k^2} |\Phi^{\dagger} \Phi|^{2}, ~~\lambda^{0} = \frac{M_{0}}{k}\Phi^{\dagger}\Phi~{\rm{and}}~~m^{2}~= ak^{2}.$$

$\lambda^{0},~\lambda^{1}$ represent the scalar potential on the UV and IR brane respectively. $M_{0},~M_{1}$ are boundary mass terms on the UV and IR brane respectively. A quartic term is added on the IR brane
to ensure electroweak symmetry breaking. $a$ represents the dimensionless bulk mass parameter defined in the units of curvature, $k$.\\
Choosing $$\Phi(x,y) = \frac{1}{\sqrt{2}}\begin{bmatrix}
           0 \\
           v(y) + H(x,y) 
         \end{bmatrix}$$ \\
and considering the metric given in Eq.~(\ref{metric}), the equation of motion for the vacuum expectation value (vev, $v(y)$) is given by (See Appendix):\\
$$\partial_{y}(e^{- 4 k y} \partial^{y} v) + e^{- 4 k y} ak^{2} v = 0\;,$$\\ with boundary conditions\\
$$ (v \partial^{y} v)_{|r \pi} = \lambda^{1}(v(y = r \pi))~~{\rm{and}}~~(v \partial^{y} v)_{|0} = 2 \lambda^{0}(v(y = 0)).$$
\\Similarly, the equation of motion for $H(x,y)$ is given by,\\
$$ e^{- 2 k y} \partial_{\mu}\partial^{\mu} H(x,y) + e^{-4 k y} ak^{2} H(x,y) + \partial_{y}(e^{-4 k y} \partial^{y} H(x,y) = 0\;,$$ \\with boundary conditions \\
$$ (H(x,y) \partial^{y} H(x,y))_{|r\pi} = \frac{\partial^{2}\lambda^{1}}{\partial H^{2}}_{|H = v} H^{2} ~~{\rm{and}}~ H(x,y) \partial^{y} H(x,y))_{|0} = \frac{\partial^{2}\lambda^{0}}{\partial H^{2}}_{|H = v}H^{2}\;.$$ \\
$H(x,y)$ is a scalar field that can be expanded in terms of its KK tower as $f_{n}^{h}h_{n}/\sqrt{\pi R}$ where $h_{n}(x)$ is the 
nth KK field with mass $m_{n}$ and $f_{n}^{h}(y)$ is the profile.
The equation of the profile $f^{h}_{n}$ is 
\begin{equation}
 -\partial_{y}(e^{-4ky} \partial_{y} f^{h}_{n}) + e^{-4ky}m^{2} f^{h}_{n} = m_{n}^{2} e^{-2ky}f^{h}_{n}\;,
\end{equation}
where $\Box h_{n} = m_{n}^{2} h_{n}.$\\

The electroweak symmetry breaking occurs on the TeV brane and the zero mode$(h_{0}~{\rm{with}}~m_{0}~=~0)$ gets its mass from the boundary potential on the TeV brane. 
Thus, the vev and the zero mode follow the same bulk profile and one can say that the 5D vev of the Higgs field is entirely carried by the zero mode. 
The potential on the UV brane is chosen such that the profile of the zero mode and the vev localises on the TeV brane and the boundary condition on the TeV brane fixes the 
mass of the Higgs with the identification $M_{1}~=~bk$. $b$ represents the dimensionless brane mass parameter in units of $k$.
Thus, we have \\
 $$\Phi(x,y) = \frac{1}{\sqrt{2 \pi R}}\begin{bmatrix}
            0 \\
            (v_{\text{SM}} + h_{0}(x))f^{h}_{0}(y) + h_{n}(x)f^{h}_{n}(y)
             \end{bmatrix}\;,$$
where\\ $$f^{h}_{0}~=~\sqrt{\frac{(2(b - 1) k \pi R)}{(e^{2(b - 1)k R \pi} - 1 )}}e^{(b-1) k y}~{\rm{and}}~b~=~2 + \sqrt{4 + a}.$$\\
 
Similarly, the bulk equation of motion of $h_{1}$ gives us the profile
$$f^{h}_{1}~=~1.85\sqrt{k R \pi} e^{-k(R \pi - y)} (J_{b -2}(\frac{m_{1} e^{ky}}{k}) + 0.36 Y_{b - 2}(\frac{m_{1} e^{ky}}{k}))\;,$$
having mass given by $m_{1}~=~(1 + 2(b-2) ) \frac{\pi}{4} k e^{- k R \pi}$. \\

\begin{figure}[h]
	\begin{center}
		\includegraphics[origin=rb,width=4.0in]{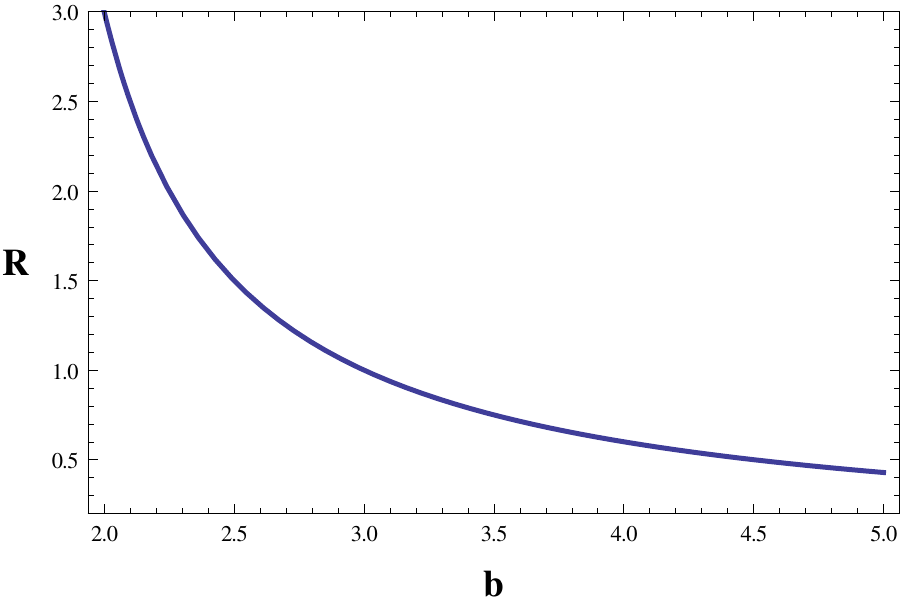}
 	\end{center}
	\caption{\it Variation of the ratio $R~=~M_{g_1}/M_{h_1}$ with b, where $M_{g_1}$ is mass of the first KK mode of gluon. } \protect\label{massratio}
\end{figure}

From figure~\ref{massratio} we see that, depending on the value of $b$, the mass of $h_1$
can be as low as the third of the first gauge boson KK mode mass. This implies
that the $h_1$ mass can be as low as a TeV in the custodial symmetry mode. When $b \gg 2$, the mass of the $h_{1}$ is heavier and can not be directly probed at the LHC.
In our analysis, we have
considered a $h_1$ with mass of 1 TeV and beyond. It is also important to note
that the best-fit point from the electroweak analysis presented in Ref.
\cite{Iyer:2015ywa} gives a value of $b~=~2$. This value of $b$ is consistent
with a $h_1$ mass of 1 TeV and, in other words, such a value of $h_1$ mass
passes the acid test of electroweak constraints. It may be noted that the normalisation
of the profile for the zero mode fixes the coupling of the SM Higgs with  all the other SM particles. Thus,
we do not expect any deviation from the observed signal strength measurement of the SM Higgs at the LHC~\cite{Aad:2015zhl,Aad:2015gba,CMS-PAS-HIG-13-005}.

The SM Higgs mixes with the radion, which is the field parametrising the fluctuation between the two branes.
In the limit of negligible back reaction, the kinetic term involving the radion and the Higgs induces the mixing~\cite{Cox:2013rva}. 
As the vev of the bulk Higgs is carried out
by the zero mode, the orthogonality
condition prevents the mixing of the first KK mode with the radion. 

One can calculate the following tree-level interaction of the KK modes with the SM particles from the action~(\ref{eqn:interact}),
\begin{itemize}
 \item $h_{1} \rightarrow V_{0}V_{0}:$ The term that governs the coupling is
 \begin{eqnarray}
   \int d^{4}x h_{1}W_{\mu}^{+} W^{\mu -} = \int d^{4} x h_{1}  W_{\mu}^{+} W^{\mu -} g_{5}^{2} \int{dy} (e^{(-4 + 2)k y} v_{\text{SM}} f^{h}_{0} f^{h}_{1}) \nonumber\\
                                          = \int d^{4} x h_{1}  W_{\mu}^{+} W^{\mu -}  g_{5}^{2} \int{dy} (e^{(-2)k y} v_{\text{SM}} f^{h}_{0} f^{h}_{1})\;.
\end{eqnarray}
The zero mode of the KK gauge bosons(i.e the $W_{\mu},Z_{\mu}$) have a flat profile and hence, the tree level coupling vanishes following the orthogonality condition of the KK profiles. 
\\
\item $h_{1} \rightarrow h_{0} h_{0}:$  Unlike the radion-$h_{1}$ mixing term, this interaction comes from the quartic scalar potential added on the TeV brane (\ref{eqn:interact}) where the trilinear coupling of the SM Higgs is
 \begin{equation}
   \int dy \delta(y - R \pi) \sqrt{-g}\left(\frac{\partial^{3} \lambda^{1}}{\partial H^{3}}_{|H = v}\right) \left(h_{0} \frac{f^{h}_{0}}{\sqrt{\pi R}}\right)^{3} \sim \int d^{4}x \lambda_{\text{SM}}v_{\text{SM}} h_{0}^{3}\;,
 \end{equation}
where $\lambda_{\text{SM}}v_{\text{SM}}~=~\dfrac{24 \gamma v(y = r\pi)}{k^{2}} f_{0}^{3}(y = R \pi).$\\
\\
The decay of the $h_{1}$ to SM Higgs is given by
\begin{equation}
 \int dy \delta(y - R \pi) \sqrt{-g}\;3\left(\frac{\partial^{3} \lambda^{1}}{\partial H^{3}}_{|H = v}\right) \left(h_{0} \frac{f^{h}_{0}}{\sqrt{\pi R}}\right)^{2} \frac{f^{h}_{1}}{\sqrt{\pi R}}h_{1} \sim \int d^{4}x \lambda_{h1hh} v_{\text{SM}} h_{0}^{2}h_{1}\;,
\end{equation}
where 
\begin{equation}
\lambda_{h1hh}~=~ = ~ 3 \frac{f_{1}}{f_{0}}_{|y = R\pi}\lambda_{\text{SM}} ~\sim~ 3.2 \lambda_{\text{SM}}.
\end{equation}
\\
 \item $h_{1} \rightarrow t_{0} t_{0}$: Relative to the Yukawa coupling term of the SM Higgs to tops, where  
\begin{equation}
  y_{5}  \int dy d^{4}x \sqrt{-g} f^{h}_{0}f^{t^{L}}_{0}f^{t^{R}}_{0} h_{0}t^{L}_{0}t^{R}_{0}\sim y_{\text{SM}}\int d^{4}x  h_{0}t^{L}_{0}t^{R}_{0}\;,
 \end{equation}
where 
$y_{5} \int dy \sqrt{-g}  f^{h}_{0}f^{t^{L}}_{0}f^{t^{R}}_{0} \sim  y_{\text{SM}}$,
the decay of $h_{1}$ to tops is given by
\\
\begin{equation}
 y_{h1t^{L}t^{R}}= y_{5} \int dy d^{4}x \sqrt{-g} f^{h}_{1}f^{t^{L}}_{0}f^{t^{R}}_{0} h_{1}t^{L}_{0}t^{R}_{0}\;.
\end{equation}
Considering the reduced normalised profiles we can write the Yukawa coupling of $h_{1}$ to fermions with respect to the SM Yukawa 
coupling as follows:-

\begin{equation}
y_{h1t^{L}t^{R}} ~=~ y_{\text{SM}} \frac{\int dy f^{h}_{1}f^{t^{L}}_{0}f^{t^{R}}_{0} }{\int dy f^{h}_{0}f^{t^{L}}_{0}f^{t^{R}}_{0}}\;.
\end{equation}
\\
For a flat 5D metric, the reduced normalised profiles for zero-mode fermions is given by
\begin{equation}
 f_{0}^{t_{L}}=\sqrt{\frac{(1-2c^{L})\pi kR}{e^{(1-2c_{L})\pi kR}-1}}e^{(\frac{1}{2}-c^{L})ky}\;,
\end{equation}
\begin{equation}
 f_{0}^{t_{R}}=\sqrt{\frac{(1+2c^{R})\pi kR}{e^{(1+2c^{R})\pi kR}-1}}e^{(\frac{1}{2}+c^{R})ky}\;,
\end{equation}

Using the $c^{L}$=0.4 and $c^{R}$=0 and $b=2$ we obtain
\begin{equation}
 y_{h1t^{L}t^{R}}=1.0755y_{\text{SM}}\;.
\end{equation}
\end{itemize}
The partial decay widths of the KK higgs to the pair of gluons, photons, tops and SM Higgs are given by,\\
\\
 $\displaystyle\Gamma(h_{1}\rightarrow gg)~=~(y_{h1t^{L}t^{R}} y_{\text{SM}})^2 \frac{M_{h_{1}}^{3}}{72\pi {v_{\text{SM}}}^{2}} \left(\frac{\alpha_{s}}{\pi}\right)^{2}\mid \Sigma_{q} I_{q}\mid^{2}\;,$\\
 \\
 $\displaystyle\Gamma(h_{1}\rightarrow \gamma \gamma)~=~(y_{h1t^{L}t^{R}} y_{\text{SM}})^2\frac{M_{h_{1}}^{3}}{16\pi {v_{\text{SM}}^{2}}} \left(\frac{\alpha}{\pi}\right)^{2}\mid \Sigma_{q} I_{q} + \Sigma_{l} I_{l}\mid^{2}\;,$\\
 \\
 $\displaystyle\Gamma(h_{1}\rightarrow h h)~=~\lambda_{h1hh}^2 \frac{\lambda_{\text{SM}}^{2} {v_{\text{SM}}}^{2}}{128 \pi M_{h_{1}}}\sqrt{1-\frac{4 m_{h}^{2}}{M_{h_{1}}^2}}\;,$\\
 \\
 $\displaystyle\Gamma(h_{1}\rightarrow f f)~=~N_{c} (y_{h1t^{L}t^{R}} y_{\text{SM}})^2  \frac{m_{f}^{2}M_{h_{1}}}{8\pi {v_{\text{SM}}}^{2}} \left(1-\frac{4 m_{f}^{2}}{M_{h_{1}}^{2}}\right)^{3/2}\;,$ where 
 $N_{c}~=~3.$\\
 \\
In the above equations, $ I_{q}~=~3[2\lambda_{q} + \lambda_{q}(4 \lambda_{q} - 1) f(\lambda_{q})]$ and $I_{l}~=~2 \lambda_{l} + \lambda_{l} (4 \lambda_{l} - 1) f(\lambda_{l}), $
  where $\lambda_{i}=\dfrac{m_{i}^{2}}{M_{h_{1}}^{2}}$, are the form factors.\\

\begin{figure}[t!]
	\begin{center}
		\includegraphics[origin=rb,width=4.0in]{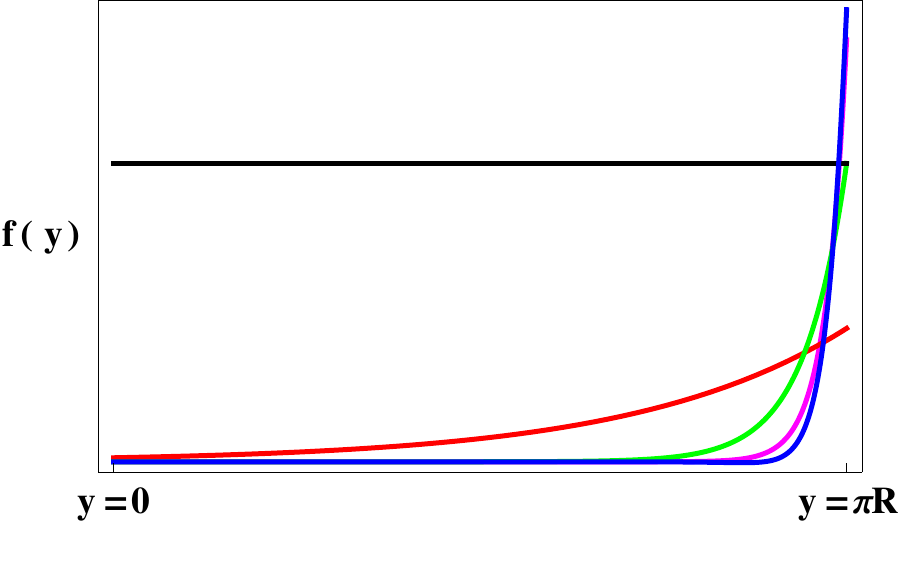}
 	\end{center}
	\caption{\it Profiles for the first KK mode of Higgs (blue), zero mode of Higgs (pink), zero mode of $t^{R}$ (green), zero mode of $t^{L}$ (red) and zero mode of
	gauge bosons (black).} \protect\label{overlap}
\end{figure}

\begin{figure}[h]
 	\begin{center}	
	\includegraphics[origin=rb,width=5.0in]{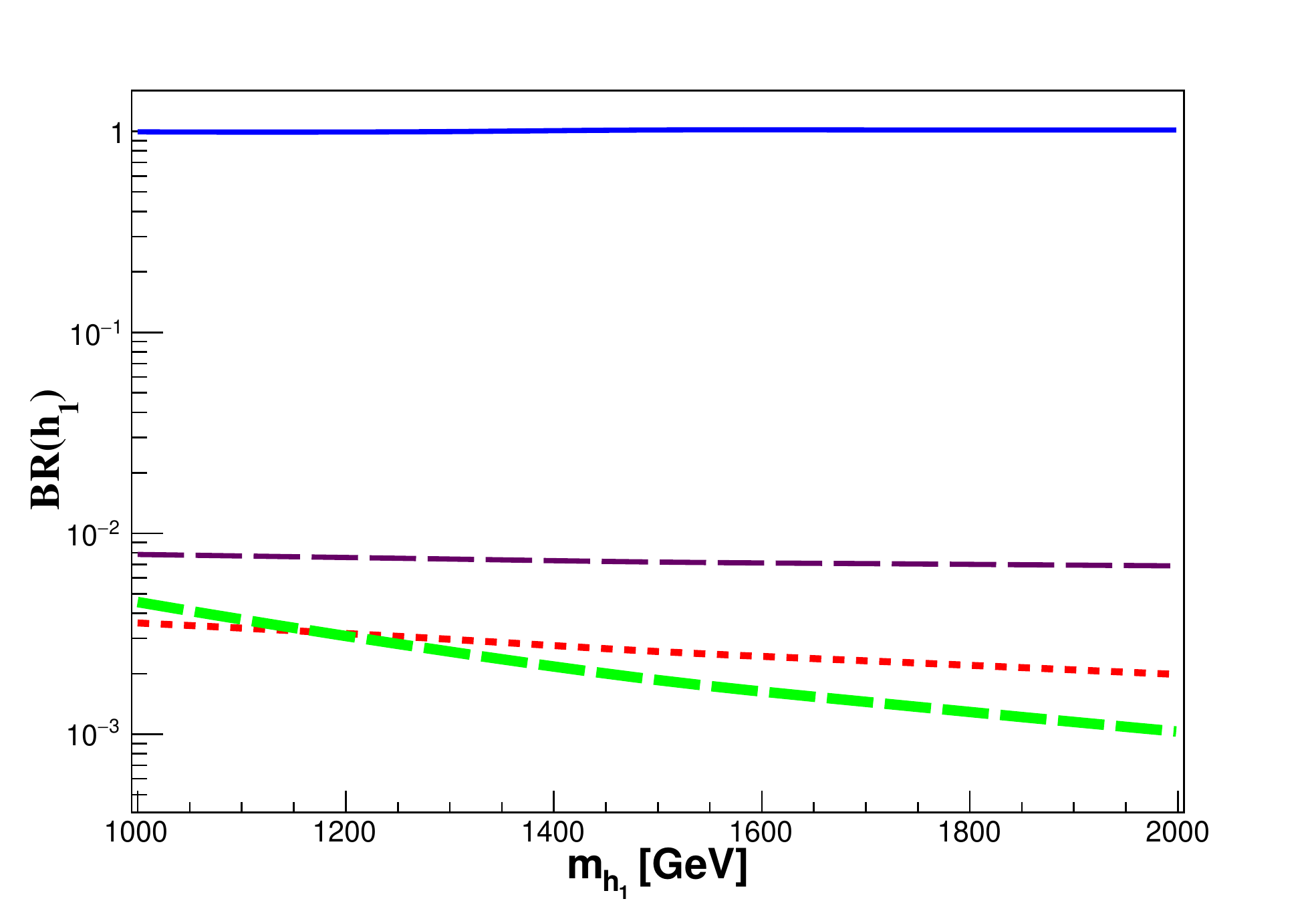}\\
 	\end{center}
 	\caption{ \it Branching ratio of the KK Higgs in $t\bar{t}$ (solid blue line), $ b\bar{b}$ (dashed magenta line), $g g$ (dashed green line) and $ h h$ (dashed red line)  channels as a function of the KK Higgs mass.}\protect\label{br_plot}
 	
\end{figure}

Having listed the couplings above for completeness, we would like to point out that the branching ratio of $h_{1}$ decaying to tops is overwhelmingly large as shown in figure~\ref{br_plot}.
Thus, we focus on the $t\bar{t}$ decay mode of $h_{1}$ in this analysis.

Since, the couplings of the KK Higgs to the massive gauge bosons vanish at the tree level, the production of the KK Higgs via vector boson fusion is heavily suppressed.
As the Yukawa coupling of the tops with the KK Higgs is of $O(1)$, the KK Higgs can be produced in association with tops or via gluon-gluon fusion with tops running in the loop.
The associated production of the KK Higgs with tops is suppressed by two orders of magnitude in this mass range. Thus, the only dominant production mode of the KK Higgs is via gluon-gluon fusion.
Even
before we launch into our analysis, we should check what constraints
existing collider data from $t \bar t$ production places on a 1 TeV
resonance decay. Recently, the ATLAS \cite{Aaboud:2016pbd} 
and the CMS \cite{Khachatryan:2015uqb}
collaborations at the LHC have presented their measurements of the
top cross-section at $\sqrt{s}=13$ TeV. The values of the cross-section
from both experiments are in agreement with NNLO QCD predictions of
the cross-section. The CMS experiment, analysing 43 pb${}^{-1}$ of data,
has quoted an error of the order of 86.5 pb on the cross-section
and the ATLAS experiment, analysing a larger 3.2 fb${}^{-1}$ sample,
has an error of the order of 36 pb. For a 1 TeV mass $h_1$, the
cross-section is much smaller (of the order of 0.5 pb). Therefore, present
measurements of the $t \bar t$ cross-section are not sensitive to
the $h_1$.

\section{$h_{1}$ at the LHC}
\begin{figure}[!t]
 	\begin{center}
 		\begin{tabular}{cc}	
 			\includegraphics[width=8 cm, height= 6cm]{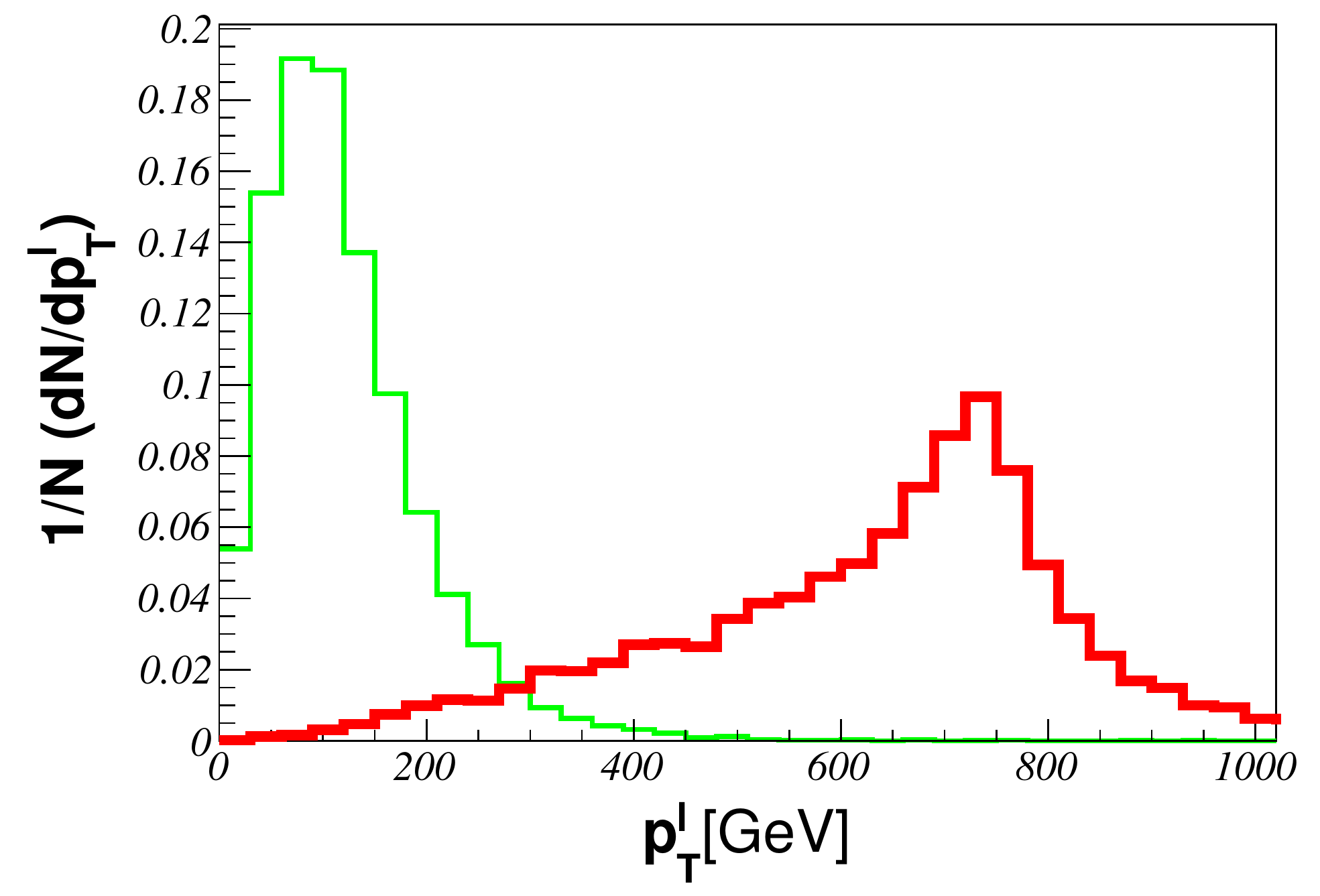}&	
 			\includegraphics[width=8 cm, height= 6cm]{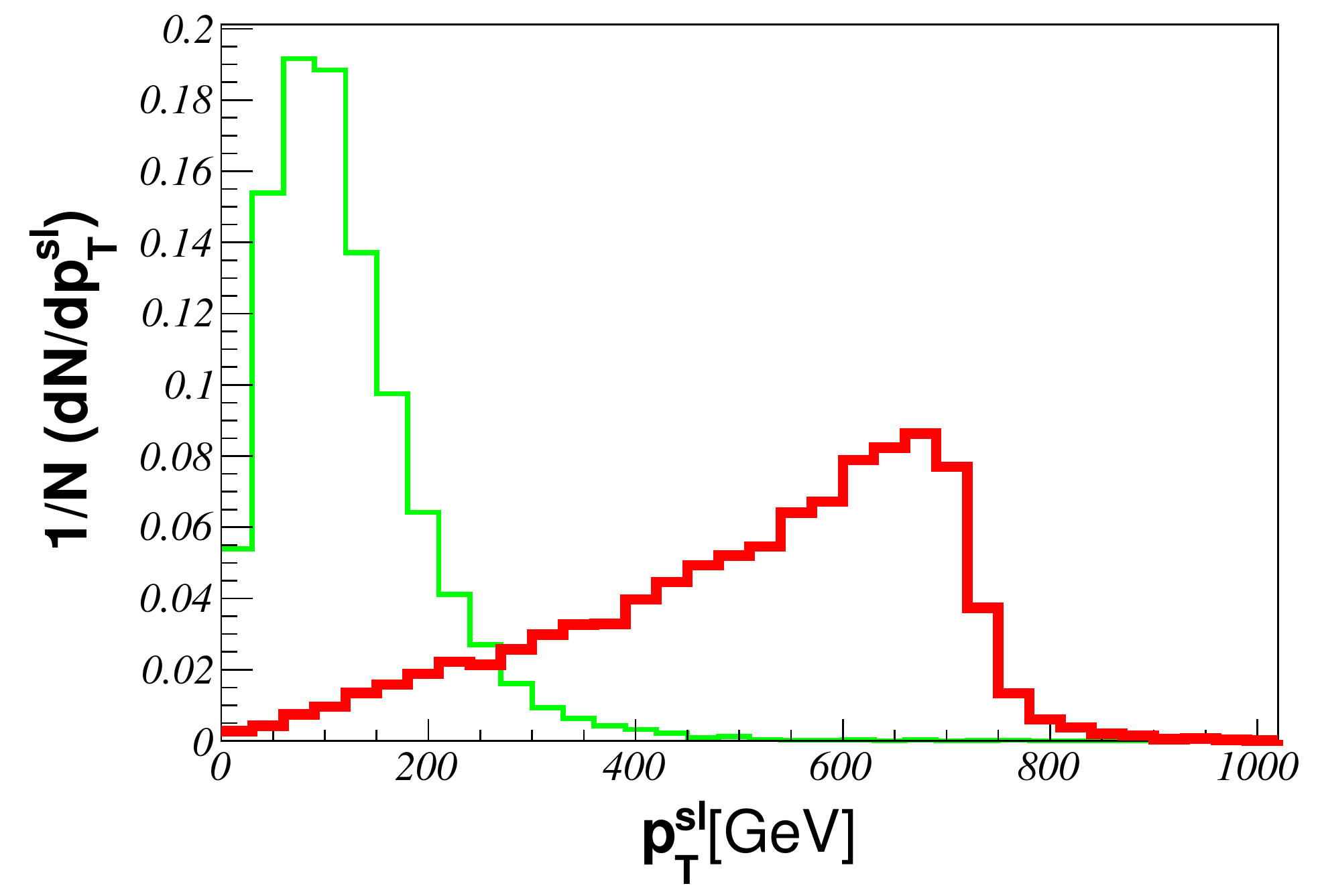}\\
 		\end{tabular}
 	\end{center}
 	\caption{\it Normalised distribution of the $p_{T}$ for leading top (left) and subleading top (right).
 	The red distribution represents signal and green represents $t\bar{t}$ background.}\protect\label{pt_top_sb}
 	
 \end{figure}

\begin{figure}
	\begin{center}
		\includegraphics[origin=rb,width=4.0in]{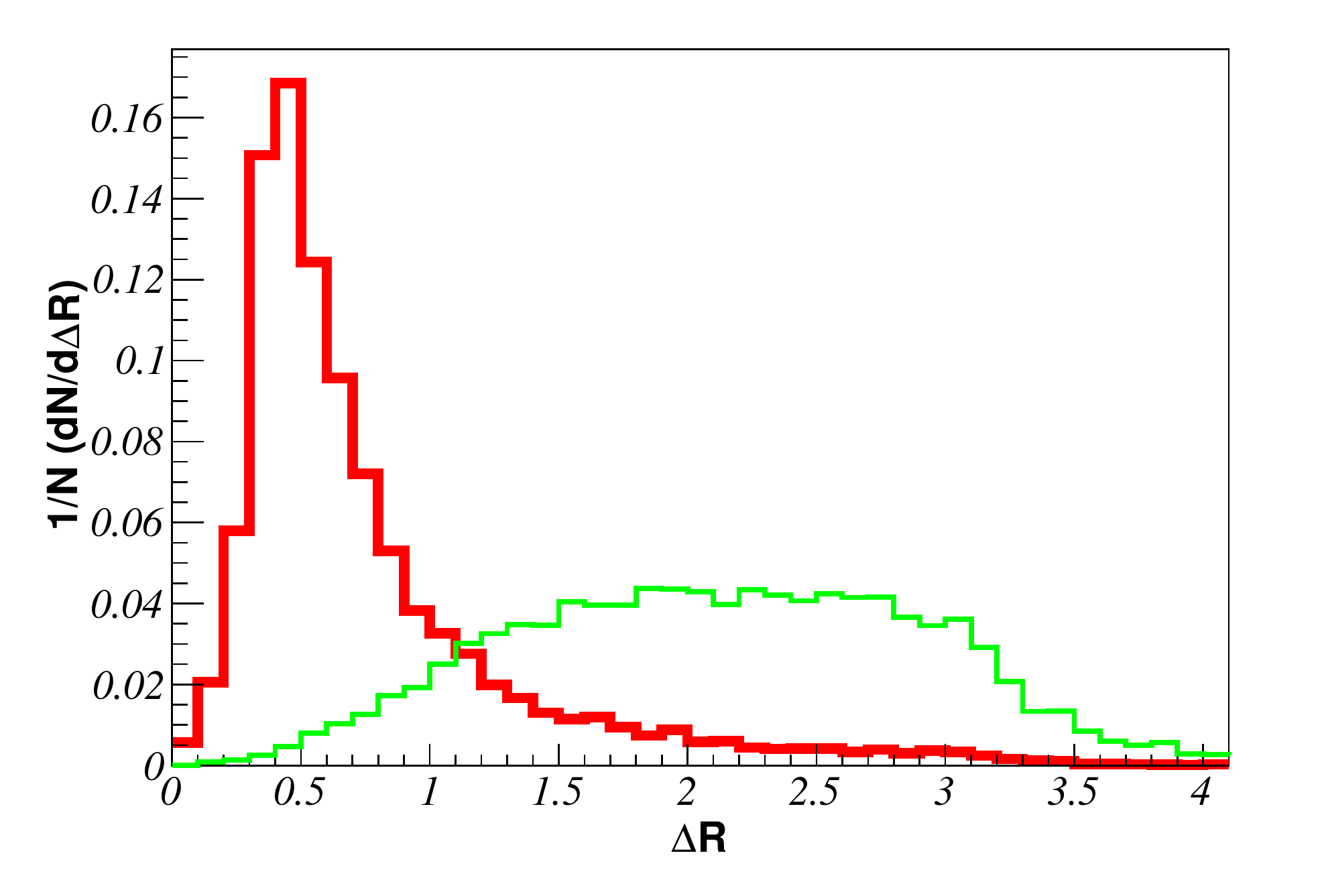}
 	\end{center}
	\caption{\it Normalised distribution of the angular separation between the decay products of the top.
	The red distribution represents signal and green represents $t\bar{t}$ background.} \protect\label{delr_wb}
\end{figure}

As discussed earlier, the $h_{1}$ is produced via gluon-gluon fusion with tops propagating in the loop and it further decays to $t\bar{t}$ at the LHC. Thus, our signal is characterised by two tops.
 Model files have been obtained with {\tt{FEYNRULES}} \cite{Alloul:2013bka}, and the signal events are generated by interfacing it with {\tt{MADGRAPH}} \cite{Alwall:2014hca} with the parton distribution function
{\tt{NNLO1}} \cite{Ball:2012cx}.
Since, we are considering the scalar having mass beyond TeV, the tops coming from the 
scalar are in the boosted regime with most of the tops having transverse momentum in the range of 200 -- 500 GeV as can be seen from the figure~\ref{pt_top_sb} and the decay products of
the top will mostly lie in a single hemisphere as can be seen in  figure~\ref{delr_wb}.

To optimize the signal, we have considered the hadronic decay of tops that
can be tagged using the HEPTopTagger \cite{Plehn:2011tg, Kasieczka:2015jma} algorithm. The backgrounds for our
signal can be categorised as
\begin{itemize}
\item Reducible background: The dominant reducible background in this topology is the dijet background. Once we demand top tagging, this background
reduces drastically. It can be controlled further using a high-transverse momentum ($p_{T}$) cut on the tagged top.  
 \item Irreducible background: The irreducible background arises from the pair production  of tops via QCD processes.
 As expected, the tagging efficiencies of the two tops are similar to the signal and hence, we need to use the decay kinematics to isolate the signal. 
\end{itemize}
The SM $t\bar{t}$ and jet events are generated using {\tt{PYTHIA 8 \cite{Sjostrand:2007gs}}}. The showering and the hadronisation of the signal event as well as the background events have been
carried out using {\tt{PYTHIA 8}}. To generate background events with larger statistics we have divided our analyses into different phase space regions
\footnote{ $\hat{p_{T}} > m_{h_{1}} - 600~{\rm GeV}~{\rm and}~~\hat{m}\in(m_{h_{1}}-300~{\rm GeV}, m_{h_{1}}+300~{\rm GeV})$ where hat represents outgoing \\parton system.}  depending upon the mass of
the KK Higgs that we are probing. 
In figure~\ref{pt_top_sb}, we have plotted the distribution of transverse momentum at the parton-level for leading (sub-leading) tops
from $h_{1}$ having mass of 1.5 TeV, SM $t\bar{t}$ background. 
As discussed earlier, the transverse momentum of tops coming from $h_{1}$ are mostly peaked near half of the $h_{1}$ mass whereas 
the SM backgrounds largely peak at the lower transverse momentum
region. Also, the decay product of the tops coming from the signal can be encompassed within a fat jet of radius $\sim 1.5$ (figure \ref{delr_wb}).
Keeping this in mind, we split our analysis into two regions. In the first region, we have reconstructed the jets using the Cambridge Aachen (C-A)
algorithm \cite{Dokshitzer:1997in,Bentvelsen:1998ug} with jet radius (R = 1.5), $p_{T}>250$ GeV 
and $|\eta|<2.4$. In the second region we have used a slightly higher value of transverse momentum to reconstruct the fat jet i.e $p_{T}>350$ GeV. 
The first part is optimised for the search of the $h_{1}$ in the range of 1 TeV whereas
the second region is proposed when its mass is around $1.5~$TeV and beyond.

These two fat jets are then considered as an input for the HEPTopTagger. The algorithm of the HEPTopTagger
is briefly described here,
\begin{itemize}
 \item Inside the fat jet one looks for hard substructure using a loose mass-drop criterion. 
 For a splitting of the fat jet $J \rightarrow j_{1}~j_{2}$, one demands that for $m_{j2} < m_{j1}, m_{j2} > 0.2 m_{J}$.
 The splitting continues till $m_{j1}<30$ GeV. The fat jets having at least 3 subjets are allowed.
 \item Once we get 3 subjets, the subjets are filtered with $R_{filt}=0.3$ and 5 filtered subjets are retained. Only those fat jets are considered which give
 total jet mass close to the top mass. These filtered
 subjets with correct top mass reconstruction are then reclustered into three subjets.
 \item These three subjets are then made to satisfy top decay kinematics. One can construct three pairs of invariant mass with these three subjets  out of which two of them are independent. In the 
 two dimensional space determined by the pair of invariant mass, top-like jets represent a thin triangular annulus (as one of them always reconstructs a W). On the other hand, the background is concentrated
 in the region of small pair-wise invariant mass.
 
\end{itemize}

\begin{figure}[h]
 	\begin{center}
 		\includegraphics[origin=rb,width=5.0in]{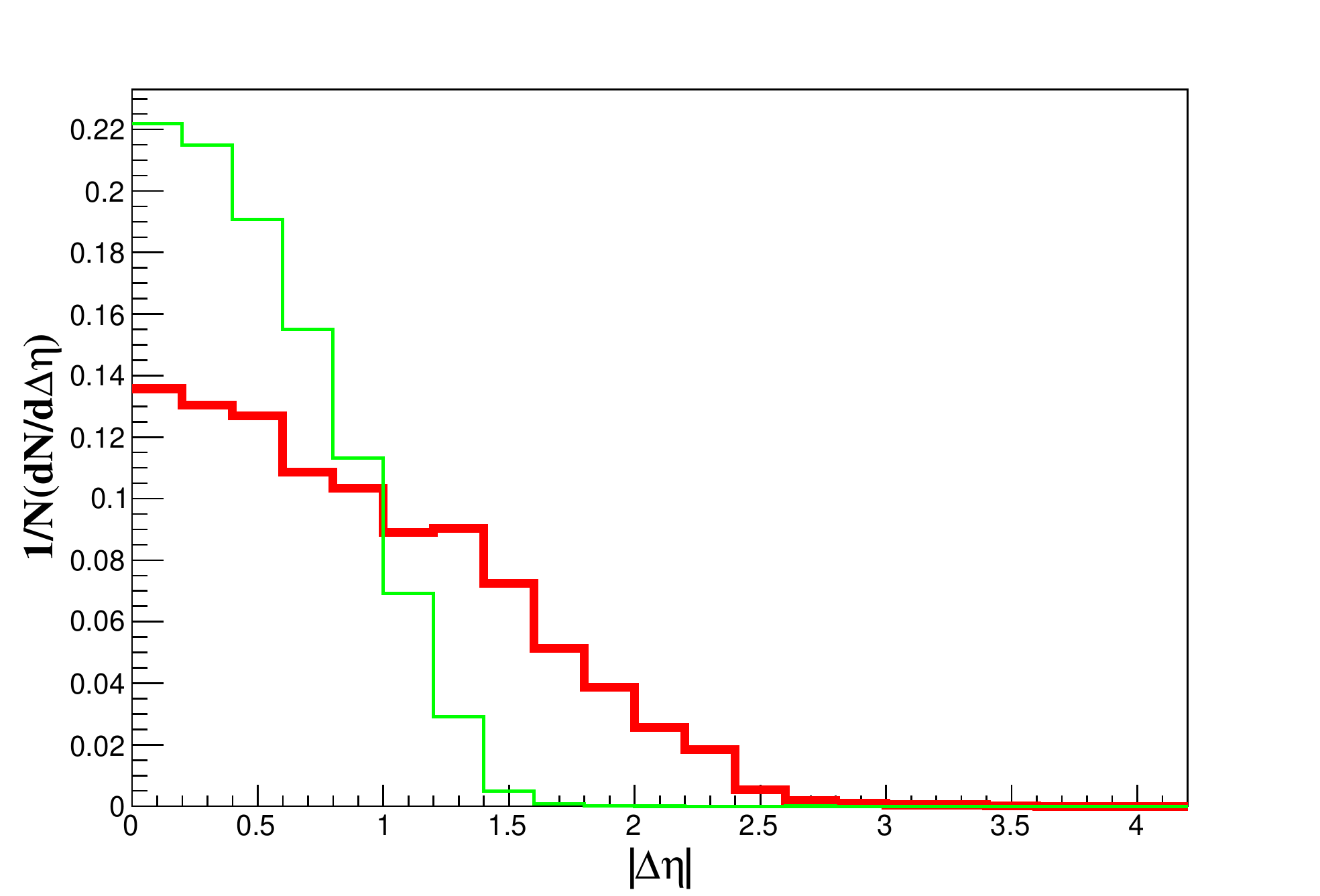}
 	\end{center}
 	\caption{\it Normalised distribution of the absolute value of $\Delta \eta$ of two top-tagged jets.
 	The red distribution represents signal for a $h_{1}$ mass of 1.5 TeV and green represents $t\bar{t}$ background.} \protect\label{fig_eta}
 \end{figure}

\begin{figure}[h]
 	\begin{center}
  		\includegraphics[origin=rb,width=5.0in]{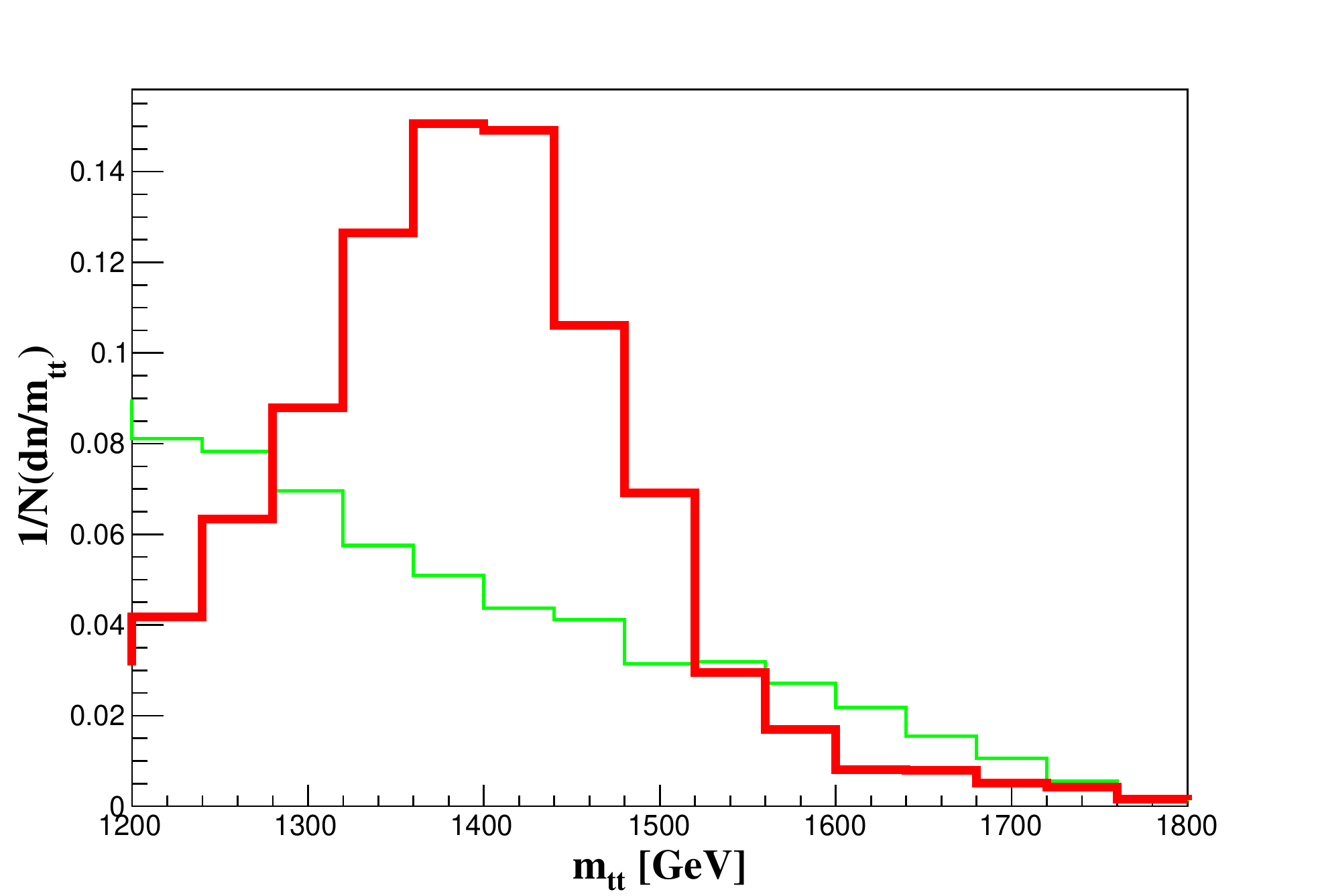}
 	\end{center}
 	\caption{\it Normalised distribution of the invariant mass of top-tagged jets pair. 
 	The red distribution represents signal for a $h_{1}$ mass of 1.5 TeV and green represents $t\bar{t}$ background.} \protect\label{fig_mtt}
 \end{figure}

We consider two such 'top-tagged' jets for our further analysis. At this stage, we have very few (almost negligible) events coming from the dijet background. The $h_{1}$ is produced
mostly at rest: as a result the top pairs are back to back. We have plotted the distribution of the absolute value of difference in rapidity $(|\Delta \eta|)$ of the 'top-tagged' pair 
coming from the signal as well as from the SM background in figure~\ref{fig_eta}. For the $t\bar{t}$ background, the distribution peaks near $|\Delta \eta|\sim 0$ whereas the tops coming from the signal
have a larger spread. We found
that a minimum cut on $\Delta \eta$ helps us to isolate signal from background. 
When the mass of the KK Higgs is around 1 TeV, we have selected events with transverse momentum of the 'top-tagged' pairs ($p_{T}^{l}$ and $p_{T}^{sl}$) greater than 350 GeV. The combination of minimum cut on
the transverse momentum and the minimum cut on pseudorapidity helps us to suppress the dijet background further. 
The efficiency of the minimum cut on $\Delta \eta$ increases as the mass of the KK Higgs increases. Thus, for the KK Higgs having a mass of 1.5 TeV and beyond, a minimum
cut on pseudorapidity is sufficient to reduce QCD as well as $t\bar{t}$.
After the angular cut, we made sure that the tops coming from the signal reconstruct the $h_{1}$ mass.
We enhance the signal efficiency by demanding that the invariant mass lies within a window about the $h_{1}$ mass.
The distribution of the invariant mass of the pair of top-tagged jets $(m_{tt})$ for the signal and $t\bar{t}$ background is plotted in
figure~\ref{fig_mtt}. Due to the effect of final state radiation (FSR), the peak of the invariant mass gets smeared mostly in the lower region of $m_{tt}$, as can be seen in figure~\ref{fig_mtt}. 
The cut flow table for two benchmark points are given in 
table~\ref{table1}. 

\begin{table}[htb]

\begin{center}
\begin{tabular}{|c|c|c|c|c|} \hline
Mass(GeV)&Cuts& Signal(fb)& QCD(fb) & $t\bar{t}$(fb) \\ \hline
1000&2 fat jets($p_{T}~>~250~{\rm{GeV}},~R~<~1.5$)& 52.36 &  395183.24 & 404.80 \\
    &2 top-tagged jets & 2.64 & 65.11 & 27.04 \\
    &$p_{T}^{l}~>~400~\rm{GeV~and~} p_{T}^{sl}~>~350~\rm{GeV}$ & 1.43 & 58.33 & 26.66 \\
    &$|\Delta \eta| > 1.15$& 0.063 & 10.39 &  1.24\\
    &$900 ~{\rm{GeV}}~<~m_{tt}~<~1100 ~{\rm{GeV}}$& 0.020 & -- & 0.005 \\\hline
1500&2 fat jets ($p_{T}~>~350~{\rm{GeV}},~R~<~1.5$)& 4.05 & 46390.00 & 91.50 \\
    &2 top-tagged jets &	0.24 & 9.24 & 5.98 \\
    &$|\Delta \eta|~>~1.3$& 0.06 & 0.41 & 0.094\\
    &$1350~{\rm{GeV}}~<~m_{tt}~<~1550~{\rm{GeV}} $&	0.04 &	-- & 0.009	\\
\hline
\end{tabular}
\end{center}
\caption{\it Cut flow table for two values of KK Higgs mass.}\label{table1}
\end{table}

\begin{figure}[t]
 	\begin{center}
  		\includegraphics[origin=rb,width=5.0in]{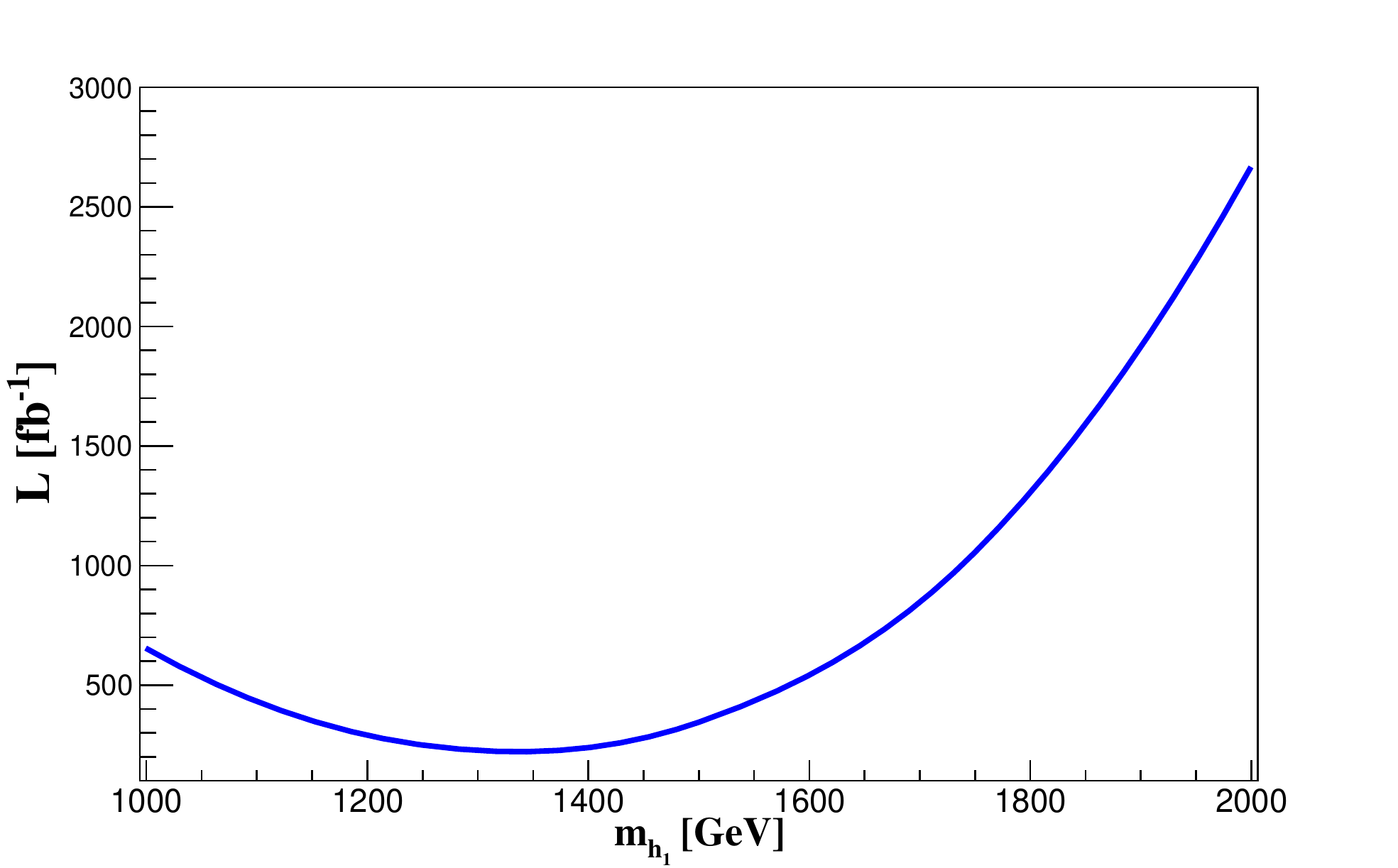}
 	\end{center}
 	\caption{\it Luminosity required to probe the KK Higgs as a function of its mass.} \protect\label{md}
 \end{figure}

Since the number of background events are comparable to the number of signal events, we calculated the significance\footnote{When $S/B \gg 1$, it coincides with our usual 
$S/\sqrt{B}.$} using\cite{Cowan:2010js},
\begin{equation}
 \sigma = \sqrt{ 2 \left(S + B\right) \log(1 + \frac{S}{B}) - 2 S}. 
 \label{significance}
\end{equation}

where $S$ is the number of signal events and $B$ is the number of background events.\\
The discovery reach for the $h_{1}$ of 1.0 TeV is about 650 fb$^{-1}$ luminosity for $\sqrt{s}~=14$ TeV.
As the mass of the KK Higgs increases, the dijet as well as $t\bar{t}$ backgrounds fall rapidly and 
one can probe it with even lower luminosity.
In figure~\ref{md}, we plotted the luminosity required to discover the KK Higgs with 5$\sigma$ discovery.
In the range of 1 TeV, due to large SM backgrounds, we applied stronger cuts which reduces the signal. Thus, we need more than 600 fb$^{-1}$ of integrated luminosity to discover it.
Once the mass increases, the SM background falls and  it is possible to observe the KK Higgs having a mass around 1.2 TeV with about 200 fb$^{-1}$ of integrated luminosity  
at $\sqrt{s}~=13$ TeV. Beyond 1.8 TeV, the production cross section decreases severely due to s-channel suppression and thus, we need about 1000 fb$^{-1}$ of integrated luminosity.

\section{Conclusion}
In order to address the gauge hierarchy problem, it is sufficient to have Higgs close to the IR brane $(b \geq 2)$ and not necessarily brane localised. 
 We find that with $b=2$, which is the best fit value consistent with the electroweak analysis, one can have $h_{1}$
 much lighter than the first KK mode of gauge bosons. The orthogonality relations among the KK profiles prevent the coupling of the $h_{1}$ to the massive gauge bosons at the tree level.
 We observe that the branching ratio of $h_{1}$ decaying to a pair of SM Higgs is about 1\%.
Thus, the $h_{1}$ decays dominantly to a pair of $t\bar{t}.$\\
We have focussued  on the $t\bar{t}$ decay mode of $h_{1}$ where both the tops are decaying hadronically. Such a $h_{1}$ is produced at the LHC via gluon-gluon fusion.
The reducible background for this topology is the SM dijet background and the irreducible background is the SM $t\bar{t}$ background.
We find that using the substructure of the boosted top, especially tagging the fat jets using HEPTopTagger,  QCD background reduces drastically.
We find that on applying cuts on the kinematic variable such as transverse momentum ($p_{T}$) and absolute value of the rapidity difference ($\Delta \eta$) of the tagged-top jets, 
we could suppress the irreducible background as well. In fact, one can discover a $h_{1}$ having a
mass lying in the range of 1.1 -1.6 TeV at 13 TeV center of mass energy with an integrated luminosity of about 300 fb$^{-1}$. The high luminosity LHC on the other hand will be able to probe the full range between 1 and 2 TeV.\\ 
\\
To conclude, we have shown that it is possible to explore the first  KK mode of Higgs hitherto considered beyond the reach of LHC.

\section{Acknowledgements}

We would like to thank Abhishek Iyer for discussion and for collaboration at the early stages of this work. 
KS was visiting CRAL, IPNL Lyon and Theory Division, CERN  while this work was carried out and would like to gratefully acknowledge their hospitality.

\appendix
\section*{Appendix A}
\setcounter{section}{1}

The action in Eq.~(\ref{action_5}) can be expressed as
 \begin{eqnarray}
  S = \int d^{4}x dy\left((-\frac{1}{2} e^{-2 ky} H \partial_{\mu}\partial^{\mu} H(x,y) - e^{-4 k y} ak^{2} \frac{H(x,y)^{2}}{2} - \frac{H(x,y)}{2} \partial_{y}(e^{-4 k y}\partial^{y} H(x,y))\right) \nonumber \\ 
    + \left(\frac{1}{2} \partial_{y} ( H e^{- 4 k y} \partial_{y} H) - \frac{\partial^{2}\lambda^{1}}{\partial H^{2}}_{|H = v} \frac{H^{2}}{2} e^{-4 k y} \delta( y - R \pi) +  \frac{\partial^{2} \lambda^{0}}{\partial H^{2}}_{|H = v}e^{-4 k y} \frac{H^{2}}{2} \delta( y)\right) \nonumber \\
    + \left(\frac{1}{2} \partial_{y} ( v e^{- 4 k y} \partial^{y} v) - \lambda^{1} e^{- 4 k y} \delta(y - R \pi) + \lambda^{0}\delta(y)\right) \nonumber \\
    + \left(\partial_{y}(H e^{- 4 k y} \partial ^{y} v) - \frac{\partial \lambda^{1}}{\partial H}_{|H = v} H e^{- 4 k y} \delta( y - R \pi) + \frac{\partial \lambda^{0}}{\partial H}_{|H = v} H e^{- 4 k y} \delta( y)\right) \nonumber \\
     + \left(\frac{-v}{2} \partial_{y}(e^{-4ky}\partial^{y} v) - \frac{e^{-4 ky}}{2} a k^{2}v^{2}\right)  + H ( -\partial_{y}(e^{-4ky} \partial^{y} v) - ak^{2} e^{-4 ky} v)) + S_{int}\;, \nonumber \\
     	\label{eqn:free}
  \end{eqnarray}
where the covariant derivative is $$ D_{M} = \partial_{M} - i g_{W} T^{a} W_{M}^{a} - i g_{B} Y B_{M}\;,$$ and 
 \begin{eqnarray}
  S_{int}  = \int d^{4} x dy \sqrt{-g}( H v (g_{W}^{2} W_{M}^{+}W^{M -} + g_{W}^{2} W_{M}^{3} W^{M 3} + g_{B}^{ 2} B_{M}B^{M} - 2g_{B}g_{W}B_{M}W^{M 3}) \nonumber\\
           +  \frac {v^{2}}{4}(g_{W}^{2} W_{M}^{+}W^{M -} + g_{W}^{2} W_{M}^{3} W^{M 3} + 2 g_{B}^{2} B_{M}B^{M} - 4 g_{B} g_{W} B_{M}W^{M 3}) \nonumber\\ 
	  +  \frac {H^{2}}{4}(g_{W}^{2} W_{M}^{+}W^{M -} + g_{W}^{2} W_{M}^{3} W^{M 3} + 2 g_{B}^{2} B_{M}B^{M} - 4 g_{B} g_{W} B_{M}W^{M 3})) \nonumber\\ 
 	  -  \left(\frac{\partial^{3} \lambda^{1}}{\partial H^{3}}_{|H = v} \frac{H^{3}}{6} +  \frac{\partial^{4} \lambda^{1}}{\partial H^{4}}_{|H = v}\frac{H^{4}}{24}\right) \delta( y - R \pi) + L_{yuk})\;. \nonumber \\
 	  \label{eqn:interact}
 \end{eqnarray}

The equation of motion for the profiles of the vev ($v(y)$) and $h^{n}(x)$ can be deduced from the expansion of the action in Eq. (\ref{eqn:free}).
The tadpole term of $H$ vanishes using equation of motion of $v(y).$

The masses of the gauge bosons are given by $M_{W}=g_{W}v_{\text{SM}}/2~,~M_{Z}=M_{W}/\cos\theta_{w}$ and $M_{\gamma}=0$
where $\cos\theta_{w}=g_{W}/\sqrt{g_{W}^{2} + g_{B}^{2}}.$

\bibliography{h1.bib}

\providecommand{\href}[2]{#2}\begingroup\raggedright\begin{thebibliography}{10}

\bibitem{Randall:1999ee}
L.~Randall and R.~Sundrum, {\it {A Large mass hierarchy from a small extra
  dimension}},  {\em Phys.Rev.Lett.} {\bf 83} (1999) 3370--3373,
  [\href{http://arxiv.org/abs/hep-ph/9905221}{{\tt hep-ph/9905221}}].

\bibitem{Gherghetta:2010cj}
T.~Gherghetta, {\it {TASI Lectures on a Holographic View of Beyond the Standard
  Model Physics}},  {\em Physics of the Large and the Small, Proceedings of the
  Theoretical Advanced Study Institute in Elementary Particle Physics, - TASI
  2009 {\rm (eds. C. Csaki and S. Dodelson)}} (2010)
  [\href{http://arxiv.org/abs/1008.2570}{{\tt arXiv:1008.2570}}].

\bibitem{Raychaudhuri:2016}
S.~Raychaudhuri and K.~Sridhar, {\em {Particle Physics of Brane Worlds and
  Extra Dimensions}}.
\newblock Cambridge University Press, 2016.

\bibitem{Pomarol:1999ad}
A.~Pomarol, {\it {Gauge bosons in a five-dimensional theory with localized
  gravity}},  {\em Phys.Lett.} {\bf B486} (2000) 153--157,
  [\href{http://arxiv.org/abs/hep-ph/9911294}{{\tt hep-ph/9911294}}].

\bibitem{Gherghetta:2000qt}
T.~Gherghetta and A.~Pomarol, {\it {Bulk fields and supersymmetry in a slice of
  AdS}},  {\em Nucl.Phys.} {\bf B586} (2000) 141--162,
  [\href{http://arxiv.org/abs/hep-ph/0003129}{{\tt hep-ph/0003129}}].

\bibitem{Grossman:1999ra}
Y.~Grossman and M.~Neubert, {\it {Neutrino masses and mixings in
  nonfactorizable geometry}},  {\em Phys.Lett.} {\bf B474} (2000) 361--371,
  [\href{http://arxiv.org/abs/hep-ph/9912408}{{\tt hep-ph/9912408}}].

\bibitem{Burdman:2003nt}
G.~Burdman, {\it {Flavor violation in warped extra dimensions and CP
  asymmetries in B decays}},  {\em Phys.Lett.} {\bf B590} (2004) 86--94,
  [\href{http://arxiv.org/abs/hep-ph/0310144}{{\tt hep-ph/0310144}}].

\bibitem{Huber:2003tu}
S.~J. Huber, {\it {Flavor violation and warped geometry}},  {\em Nucl.Phys.}
  {\bf B666} (2003) 269--288, [\href{http://arxiv.org/abs/hep-ph/0303183}{{\tt
  hep-ph/0303183}}].

\bibitem{Casagrande:2008hr}
S.~Casagrande, F.~Goertz, U.~Haisch, M.~Neubert, and T.~Pfoh, {\it {Flavor
  Physics in the Randall-Sundrum Model: I. Theoretical Setup and Electroweak
  Precision Tests}},  {\em JHEP} {\bf 0810} (2008) 094,
  [\href{http://arxiv.org/abs/0807.4937}{{\tt arXiv:0807.4937}}].

\bibitem{Bauer:2009cf}
M.~Bauer, S.~Casagrande, U.~Haisch, and M.~Neubert, {\it {Flavor Physics in the
  Randall-Sundrum Model: II. Tree-Level Weak-Interaction Processes}},  {\em
  JHEP} {\bf 1009} (2010) 017, [\href{http://arxiv.org/abs/0912.1625}{{\tt
  arXiv:0912.1625}}].

\bibitem{Agashe:2004cp}
K.~Agashe, G.~Perez, and A.~Soni, {\it {Flavor structure of warped extra
  dimension models}},  {\em Phys.Rev.} {\bf D71} (2005) 016002,
  [\href{http://arxiv.org/abs/hep-ph/0408134}{{\tt hep-ph/0408134}}].

\bibitem{Quiros:2013yaa}
M.~Quiros, {\it {Higgs Bosons in Extra Dimensions}},  {\em Mod. Phys. Lett.}
  {\bf A30} (2015), no.~15 1540012, [\href{http://arxiv.org/abs/1311.2824}{{\tt
  arXiv:1311.2824}}].

\bibitem{Agashe:2003zs}
K.~Agashe, A.~Delgado, M.~J. May, and R.~Sundrum, {\it {RS1, custodial isospin
  and precision tests}},  {\em JHEP} {\bf 0308} (2003) 050,
  [\href{http://arxiv.org/abs/hep-ph/0308036}{{\tt hep-ph/0308036}}].

\bibitem{Agashe:2006at}
K.~Agashe, R.~Contino, L.~Da~Rold, and A.~Pomarol, {\it {A Custodial symmetry
  for Zb anti-b}},  {\em Phys.Lett.} {\bf B641} (2006) 62--66,
  [\href{http://arxiv.org/abs/hep-ph/0605341}{{\tt hep-ph/0605341}}].

\bibitem{Davoudiasl:2009cd}
H.~Davoudiasl, S.~Gopalakrishna, E.~Ponton, and J.~Santiago, {\it {Warped
  5-Dimensional Models: Phenomenological Status and Experimental Prospects}},
  {\em New J. Phys.} {\bf 12} (2010) 075011,
  [\href{http://arxiv.org/abs/0908.1968}{{\tt arXiv:0908.1968}}].

\bibitem{Iyer:2015ywa}
A.~M. Iyer, K.~Sridhar, and S.~K. Vempati, {\it {Bulk Randall-Sundrum models,
  electroweak precision tests, and the 125 GeV Higgs}},  {\em Phys. Rev.} {\bf
  D93} (2016), no.~7 075008, [\href{http://arxiv.org/abs/1502.06206}{{\tt
  arXiv:1502.06206}}].

\bibitem{Cabrer:2010si}
J.~A. Cabrer, G.~von Gersdorff, and M.~Quiros, {\it {Warped Electroweak
  Breaking Without Custodial Symmetry}},  {\em Phys.Lett.} {\bf B697} (2011)
  208--214, [\href{http://arxiv.org/abs/1011.2205}{{\tt arXiv:1011.2205}}].

\bibitem{Cabrer:2011fb}
J.~A. Cabrer, G.~von Gersdorff, and M.~Quiros, {\it {Suppressing Electroweak
  Precision Observables in 5D Warped Models}},  {\em JHEP} {\bf 05} (2011) 083,
  [\href{http://arxiv.org/abs/1103.1388}{{\tt arXiv:1103.1388}}].

\bibitem{Carena:2003fx}
M.~Carena, A.~Delgado, E.~Ponton, T.~M.~P. Tait, and C.~E.~M. Wagner, {\it
  {Precision electroweak data and unification of couplings in warped extra
  dimensions}},  {\em Phys. Rev.} {\bf D68} (2003) 035010,
  [\href{http://arxiv.org/abs/hep-ph/0305188}{{\tt hep-ph/0305188}}].

\bibitem{Agashe:2006hk}
K.~Agashe, A.~Belyaev, T.~Krupovnickas, G.~Perez, and J.~Virzi, {\it {LHC
  Signals from Warped Extra Dimensions}},  {\em Phys.Rev.} {\bf D77} (2008)
  015003, [\href{http://arxiv.org/abs/hep-ph/0612015}{{\tt hep-ph/0612015}}].

\bibitem{Lillie:2007yh}
B.~Lillie, L.~Randall, and L.-T. Wang, {\it {The Bulk RS KK-gluon at the LHC}},
   {\em JHEP} {\bf 0709} (2007) 074,
  [\href{http://arxiv.org/abs/hep-ph/0701166}{{\tt hep-ph/0701166}}].

\bibitem{Guchait:2007jd}
M.~Guchait, F.~Mahmoudi, and K.~Sridhar, {\it {Associated production of a
  Kaluza-Klein excitation of a gluon with a t anti-t pair at the LHC}},  {\em
  Phys.Lett.} {\bf B666} (2008) 347--351,
  [\href{http://arxiv.org/abs/0710.2234}{{\tt arXiv:0710.2234}}].

\bibitem{Allanach:2009vz}
B.~C. Allanach, F.~Mahmoudi, J.~P. Skittrall, and K.~Sridhar, {\it
  {Gluon-initiated production of a Kaluza-Klein gluon in a Bulk Randall-Sundrum
  model}},  {\em JHEP} {\bf 1003} (2010) 014,
  [\href{http://arxiv.org/abs/0910.1350}{{\tt arXiv:0910.1350}}].

\bibitem{Agashe:2007ki}
K.~Agashe, H.~Davoudiasl, S.~Gopalakrishna, T.~Han, G.-Y. Huang, et~al., {\it
  {LHC Signals for Warped Electroweak Neutral Gauge Bosons}},  {\em Phys.Rev.}
  {\bf D76} (2007) 115015, [\href{http://arxiv.org/abs/0709.0007}{{\tt
  arXiv:0709.0007}}].

\bibitem{Agashe:2008jb}
K.~Agashe, S.~Gopalakrishna, T.~Han, G.-Y. Huang, and A.~Soni, {\it {LHC
  Signals for Warped Electroweak Charged Gauge Bosons}},  {\em Phys.Rev.} {\bf
  D80} (2009) 075007, [\href{http://arxiv.org/abs/0810.1497}{{\tt
  arXiv:0810.1497}}].

\bibitem{Iyer:2016yjb}
A.~M. Iyer, F.~Mahmoudi, N.~Manglani, and K.~Sridhar, {\it {Kaluza-Klein gluon
  + jets associated production at the Large Hadron Collider}},  {\em Phys.
  Lett.} {\bf B759} (2016) 342--348,
  [\href{http://arxiv.org/abs/1601.02033}{{\tt arXiv:1601.02033}}].

\bibitem{Agashe:2004ci}
K.~Agashe and G.~Servant, {\it {Warped unification, proton stability and dark
  matter}},  {\em Phys. Rev. Lett.} {\bf 93} (2004) 231805,
  [\href{http://arxiv.org/abs/hep-ph/0403143}{{\tt hep-ph/0403143}}].

\bibitem{Davoudiasl:2007wf}
H.~Davoudiasl, T.~G. Rizzo, and A.~Soni, {\it {On direct verification of warped
  hierarchy-and-flavor models}},  {\em Phys. Rev.} {\bf D77} (2008) 036001,
  [\href{http://arxiv.org/abs/0710.2078}{{\tt arXiv:0710.2078}}].

\bibitem{Davoudiasl:2005uu}
H.~Davoudiasl, B.~Lillie, and T.~G. Rizzo, {\it {Off-the-wall Higgs in the
  universal Randall-Sundrum model}},  {\em JHEP} {\bf 08} (2006) 042,
  [\href{http://arxiv.org/abs/hep-ph/0508279}{{\tt hep-ph/0508279}}].

\bibitem{Cacciapaglia:2006mz}
G.~Cacciapaglia, C.~Csaki, G.~Marandella, and J.~Terning, {\it {The Gaugephobic
  Higgs}},  {\em JHEP} {\bf 02} (2007) 036,
  [\href{http://arxiv.org/abs/hep-ph/0611358}{{\tt hep-ph/0611358}}].

\bibitem{Frank:2016vtv}
M.~Frank, N.~Pourtolami, and M.~Toharia, {\it {Bulk Higgs and the 750 GeV
  diphoton signal}},  \href{http://arxiv.org/abs/1607.04534}{{\tt
  arXiv:1607.04534}}.

\bibitem{Aad:2015zhl}
{\bf ATLAS, CMS} Collaboration, G.~Aad et~al., {\it {Combined Measurement of
  the Higgs Boson Mass in $pp$ Collisions at $\sqrt{s}=7$ and 8 TeV with the
  ATLAS and CMS Experiments}},  {\em Phys. Rev. Lett.} {\bf 114} (2015) 191803,
  [\href{http://arxiv.org/abs/1503.07589}{{\tt arXiv:1503.07589}}].

\bibitem{Aad:2015gba}
{\bf ATLAS} Collaboration, G.~Aad et~al., {\it {Measurements of the Higgs boson
  production and decay rates and coupling strengths using pp collision data at
  $\sqrt{s}=7$ and 8 TeV in the ATLAS experiment}},  {\em Eur. Phys. J.} {\bf
  C76} (2016), no.~1 6, [\href{http://arxiv.org/abs/1507.04548}{{\tt
  arXiv:1507.04548}}].

\bibitem{CMS-PAS-HIG-13-005}
{\bf CMS} Collaboration, {\it {Combination of standard model Higgs boson
  searches and measurements of the properties of the new boson with a mass near
  125 GeV}},  Tech. Rep. CMS-PAS-HIG-13-005, CERN, Geneva, 2013.

\bibitem{Cox:2013rva}
P.~Cox, A.~D. Medina, T.~S. Ray, and A.~Spray, {\it {Radion/Dilaton-Higgs
  Mixing Phenomenology in Light of the LHC}},  {\em JHEP} {\bf 02} (2014) 032,
  [\href{http://arxiv.org/abs/1311.3663}{{\tt arXiv:1311.3663}}].

\bibitem{Aaboud:2016pbd}
{\bf ATLAS} Collaboration, M.~Aaboud et~al., {\it {Measurement of the
  $t\bar{t}$ production cross-section using $e\mu$ events with b-tagged jets in
  pp collisions at $\sqrt{s}$=13 TeV with the ATLAS detector}},
  \href{http://arxiv.org/abs/1606.02699}{{\tt arXiv:1606.02699}}.

\bibitem{Khachatryan:2015uqb}
{\bf CMS} Collaboration, V.~Khachatryan et~al., {\it {Measurement of the top
  quark pair production cross section in proton-proton collisions at $\sqrt(s)
  =$ 13 TeV}},  {\em Phys. Rev. Lett.} {\bf 116} (2016), no.~5 052002,
  [\href{http://arxiv.org/abs/1510.05302}{{\tt arXiv:1510.05302}}].

\bibitem{Alloul:2013bka}
A.~Alloul, N.~D. Christensen, C.~Degrande, C.~Duhr, and B.~Fuks, {\it
  {FeynRules 2.0 - A complete toolbox for tree-level phenomenology}},  {\em
  Comput. Phys. Commun.} {\bf 185} (2014) 2250--2300,
  [\href{http://arxiv.org/abs/1310.1921}{{\tt arXiv:1310.1921}}].

\bibitem{Alwall:2014hca}
J.~Alwall, R.~Frederix, S.~Frixione, V.~Hirschi, F.~Maltoni, O.~Mattelaer,
  H.~S. Shao, T.~Stelzer, P.~Torrielli, and M.~Zaro, {\it {The automated
  computation of tree-level and next-to-leading order differential cross
  sections, and their matching to parton shower simulations}},  {\em JHEP} {\bf
  07} (2014) 079, [\href{http://arxiv.org/abs/1405.0301}{{\tt
  arXiv:1405.0301}}].

\bibitem{Ball:2012cx}
R.~D. Ball et~al., {\it {Parton distributions with LHC data}},  {\em Nucl.
  Phys.} {\bf B867} (2013) 244--289,
  [\href{http://arxiv.org/abs/1207.1303}{{\tt arXiv:1207.1303}}].

\bibitem{Plehn:2011tg}
T.~Plehn and M.~Spannowsky, {\it {Top Tagging}},  {\em J. Phys.} {\bf G39}
  (2012) 083001, [\href{http://arxiv.org/abs/1112.4441}{{\tt
  arXiv:1112.4441}}].

\bibitem{Kasieczka:2015jma}
G.~Kasieczka, T.~Plehn, T.~Schell, T.~Strebler, and G.~P. Salam, {\it
  {Resonance Searches with an Updated Top Tagger}},  {\em JHEP} {\bf 06} (2015)
  203, [\href{http://arxiv.org/abs/1503.05921}{{\tt arXiv:1503.05921}}].

\bibitem{Sjostrand:2007gs}
T.~Sjostrand, S.~Mrenna, and P.~Z. Skands, {\it {A Brief Introduction to PYTHIA
  8.1}},  {\em Comput. Phys. Commun.} {\bf 178} (2008) 852--867,
  [\href{http://arxiv.org/abs/0710.3820}{{\tt arXiv:0710.3820}}].

\bibitem{Dokshitzer:1997in}
Y.~L. Dokshitzer, G.~D. Leder, S.~Moretti, and B.~R. Webber, {\it {Better jet
  clustering algorithms}},  {\em JHEP} {\bf 08} (1997) 001,
  [\href{http://arxiv.org/abs/hep-ph/9707323}{{\tt hep-ph/9707323}}].

\bibitem{Bentvelsen:1998ug}
S.~Bentvelsen and I.~Meyer, {\it {The Cambridge jet algorithm: Features and
  applications}},  {\em Eur. Phys. J.} {\bf C4} (1998) 623--629,
  [\href{http://arxiv.org/abs/hep-ph/9803322}{{\tt hep-ph/9803322}}].

\bibitem{Cowan:2010js}
G.~Cowan, K.~Cranmer, E.~Gross, and O.~Vitells, {\it {Asymptotic formulae for
  likelihood-based tests of new physics}},  {\em Eur. Phys. J.} {\bf C71}
  (2011) 1554, [\href{http://arxiv.org/abs/1007.1727}{{\tt arXiv:1007.1727}}].
  [Erratum: Eur. Phys. J.C73,2501(2013)].

\end{thebibliography}\endgroup
\end{document}